\documentstyle[preprint,aps,epsf,floats]{revtex}
\input epsf
\tighten       

\def\wt{\widetilde}
\def\wh{\widehat}

\def\Cs#1#2{C_{#1}^{[#2]}}
\def\si{{}^1\kern-.14em S_0}
\def\siii{{}^3\kern-.14em S_1}
\def\diii{{}^3\kern-.14em D_1}

\def\bfq{{\bf q}}

\def\MeV{{\rm\ MeV}}
\def\GeV{{\rm\ GeV}}
\def\Lnn{\Lambda_{\rm NN}}
\def\order{{\cal O}}

\def\ket#1{\vert#1\rangle}
\def\bra#1{\langle#1\vert}
\def\ltap{\ \raise.3ex\hbox{$<$\kern-.75em\lower1ex\hbox{$\sim$}}\ }
\def\gtap{\ \raise.3ex\hbox{$>$\kern-.75em\lower1ex\hbox{$\sim$}}\ }
\def\CA{{\cal A}}
\def\){\right)}
\def\({\left( }
\def\]{\right] }
\def\[{\left[ }
\def\si{{}^1\kern-.14em S_0}
\def\siii{{}^3\kern-.14em S_1}

\newcommand{\eqn}[1]{\label{eq:#1}}
\newcommand{\refeq}[1]{(\ref{eq:#1})}
\newcommand{\fig}[1]{Fig.~\ref{fig:#1}}
\newcommand{\eq}{Eq.~\refeq}

\newcommand{\eqsii}[2]{Eqs.~(\ref{eq:#1}, \ref{eq:#2})}
\newcommand{\beq}{\begin{eqnarray}}
\newcommand{\eeq}{\end{eqnarray}}

\def\Journal#1#2#3#4{{#1} {\bf #2}, #3 (#4)}

\def\NPB{Nucl.\ Phys.\ B}
\def\NPA{Nucl.\ Phys.\ A}
\def\PLB{Phys.\ Lett.\ B}
\def\PRL{Phys.\ Rev.\ Lett.}

\def\PRC{Phys.\ Rev.\ C}
\def\PRA{Phys.\ Rev.\ A}
\def\PR{Phys.\ Rev.}
\def\AP{Ann.\ Phys.}

\begin{document}
\preprint{\vbox{
\hbox{CERN-TH/99-138}
\hbox{DOE/ER/40561-55-INT99}
\hbox{SCIPP/99/21}
}}
\vskip1truein
\title{The Long and Short of Nuclear Effective Field Theory Expansions}
\author{David B. Kaplan \footnote{Address until June 30, 1999: Division 
    TH, CERN, CH1211 Geneva 23, Switzerland.}}
\address{
Institute for Nuclear Theory 351550, University of Washington, Seattle, WA
98195--1550, USA  \\  {\tt
dbkaplan@phys.washington.edu} }
\author{James V.  Steele }
\address{
Department of Physics, The Ohio State University,  
Columbus, OH 43210--1106, USA  \\  {\tt
jsteele@mps.ohio-state.edu}  }
\maketitle

\begin{abstract} 

Nonperturbative effective field theory calculations 
for $NN$ scattering seem to break down at rather low momenta. 
By examining several
toy models, we clarify how effective field theory
expansions can in general be used to properly separate long- and 
short-range effects.
We find that one-pion exchange has a large effect on the
scattering phase shift near poles in the amplitude, but otherwise can
be treated  perturbatively.
Analysis of a toy model that reproduces $\si$ $NN$ scattering data
rather well suggests that failures of effective field theories 
for momenta above the pion mass can be due to
short-range physics rather than the treatment of pion exchange. We
discuss the implications this has for extending the applicability of 
effective field theories.

\end{abstract}

\thispagestyle{empty}

\bigskip
\vfill
\leftline{May 14, 1999}
\eject
\section{Introduction}
\label{sec:1}

Effective field theories (EFTs) allow for a description of low-energy data
to arbitrary accuracy
without detailed knowledge of the short-range interactions. 
The technique is most successful when there is a clear hierarchy of
scales, for then the short-range physics only affects low-energy
processes through weak irrelevant operators suppressed by powers of
the short-distance scale.  
A good example is the $SU(3)\times SU(2)\times U(1)$ 
gauge theory for elementary particles, which is thought to be the
effective low-energy formulation of a more complete theory.
Experimental lower bounds on the momentum scale 
characterizing particular irrelevant operators from physics beyond the
standard model are as high as  
$10^{16}\GeV$.
In contrast, protein folding, for example, probably cannot benefit
from an EFT  
analysis due to the absence of a large gap in the energy spectrum.
Chiral perturbation theory is a successful EFT because the
pion mass is small compared to other hadrons, as guaranteed in the
chiral symmetry limit of QCD.   
Having an approximate symmetry to produce a small expansion parameter
is important,
because symmetry arguments can be used to unambiguously determine the
importance of any particular process by counting powers of the small
parameter.

Recently, there has been
much interest in applying chiral perturbation theory to nuclear
physics \cite{all}.
For low enough momentum, only interactions between pions and nucleons
are relevant, with
the effects of heavier hadrons (such as vector mesons and delta baryons)
accounted for by higher dimension contact interactions.  
The primary goal is to uncover a systematic expansion of 
nuclear forces at low energy in powers of the pion mass and external
momenta divided by a typical QCD scale $\Lambda_S$.  
This scale $\Lambda_S$ should be somewhat larger than $m_\pi$ for the
expansion to 
be useful, but is unknown {\it a priori}.  

A complication in nucleon dynamics that does not occur for chiral
perturbation theory with zero or one nucleon
is the appearance of a new long
length scale, the nucleon-nucleon scattering length.  
The large value of the scattering length apparently arises
from a fortuitous fine tuning of the short-distance physics.
The situation is analogous to a condensed matter system
near a phase transition, where interactions 
take on critical strengths, resulting in correlation lengths much
longer than the lattice spacing.  
Similarly, the EFT for nucleons appears to
be near a nontrivial critical point, and therefore the scaling dimensions of
operators can be quite different from their naive dimension. This must
indeed be the case, as all $NN$ interactions are  irrelevant by
naive dimensional analysis, yet they are quite strong in
reality. 

Weinberg's seminal papers \cite{Weinberg1} and their subsequent
application  \cite{KoMany,Parka,steelea} outline a way to
apply EFT methods to nuclear physics.
In the spirit of the
Hamiltonian theory of Wilson with additional constraints imposed 
by chiral symmetry  (in particular, the fact that the pion is
derivatively coupled), the kernel of the Lippmann-Schwinger
equation (i.e.\ the effective potential) is expanded in powers of
$p/\Lambda_S$ and $m_\pi/\Lambda_S$  and then 
iterated to all orders.  
The scale $\Lambda_S$ is assumed not to vanish in the chiral limit and
can be determined from fitting to data.  
In this approach, the way a given operator depends on $\Lambda_S$
is determined simply by dimensional analysis. In
particular, the leading effects in this approach are due to one pion exchange
(OPE) and a four nucleon contact interaction $(N^\dagger N)^2$,
which both scale as $1/\Lambda_S^2$. 

However, since the interactions are tuned to a critical point,
true and naive dimensions of operators differ and it 
is far from obvious that Weinberg's power counting is 
correct~\cite{Kaplan:1998hb,birse1}.
An alternative approach was therefore proposed by Kaplan, Savage, and
Wise (KSW) \cite{KSW}, involving an
expansion about the critical point corresponding to an infinite $s$-wave
scattering length. In this analysis, one finds, for example, that only the 
$ (N^\dagger N)^2$ operator is marginal, while OPE and other
interactions are irrelevant and subleading. 
The effects of the leading contact interaction can be summed analytically,
and the amplitude is expanded perturbatively in 
the higher dimension operators.

As we discuss below, the difference between the
Weinberg and KSW expansions 
for $NN$ scattering at any given order only differ by
terms considered higher order in the KSW approach, provided the
renormalization scale in the former method is chosen to be
sufficiently high.  It is not clear that this correspondence will be
true for systems with more than two nucleons.
In any case, the KSW approach allows for a potentially great simplification
compared to Weinberg's proposal, in particular for the case of many nucleon 
systems where iterating extended interactions to all orders is
numerically intensive. 
 
There has been much discussion and some controversy in the literature
about which expansion scheme to use for an effective field theory of
nuclear forces, and whether either one works at all \cite{cohena}.
Both schemes have had some success in the 
two- and three-nucleon sector~\cite{all}, but it is not clear that they 
constitute  an improvement over effective range theory~\cite{bethe},
or the related  
(but more sophisticated) pionless nuclear EFT \cite{CRS}.

We address here the nature and implementation of the KSW
expansion in the context of simple quantum mechanical models of the
$NN$ interaction. 
We begin by analyzing a system with only short-range interactions and 
contrast the Weinberg and KSW expansions.  We then
consider arbitrary short-range interactions in conjunction with a
caricature of single pion exchange, modeled by
a delta-shell potential with radius $1/m_\pi$.  
This model serves to illustrate how the effective field theory
works when both long and short length scales are present. Based on our 
analytical treatment of this toy model, we develop a general method
for computing the low-energy constants of the effective field theory.
We then extend our analysis to more realistic examples 
and conclude with a discussion of the status of the KSW expansion.

\section{Short-range  Interactions}
\label{sec:2}

We assume that interactions between two particles we will call
``nucleons'' can be 
described exactly by a potential $V=V_S+V_L$ consisting of a short-
and long-range part.
We simplify things even further in this section by taking the
long-range potential $V_L$ to be zero.
Within the context of field theory, it is
most natural to express the scattering problem in terms of the Feynman
scattering amplitude.  
Focusing exclusively on the $\si$ partial wave below, the amplitude is
given by 
\beq
\CA \equiv {4\pi\over M}\({e^{2 i\wt \delta} -1\over 2 i p}\) =
{4\pi/M\over p\cot\wt\delta - i p}\ ,
\eqn{ampdef}
\eeq
where $p=\sqrt{M E_{cm}}$ is the center of mass momentum of the two
nucleons and  $\wt\delta$ is the phase
shift for the short-range potential $V_S$ alone.

A successful EFT should describe the scattering amplitude in terms of
an expansion in powers of momentum over the characteristic
short-distance scale $\Lambda_S$.  
It was already evident fifty years ago that such an expansion is
possible with the discovery of the effective range expansion
\cite{bethe}, which expresses the quantity $p\cot\wt\delta$ as
\beq
p\cot\wt\delta = -\frac{1 }{\wt a}+ \frac12 \wt r_0 p^2+
\sum_{n=2}^\infty\wt v_n p^{2n}\ ,
\eqn{ere}
\eeq
where $\wt a$ is the scattering length, $\wt r_0$ is the
effective range, and the $\wt v_n$ are higher terms in the Taylor
expansion.
All the effective range parameters are expected to scale with the 
appropriate power of $\Lambda_S$ with the exception of 
$\wt a$, which can take on any value.
We will focus on the case where the scattering length is abnormally
large $|\wt a| \gg 1/\Lambda_S$ (as occurs in actual $NN$ scattering), 
which signals the presence of a state that is weakly bound or almost
bound.

\subsection{The Exact Solution of the Effective Theory}
\label{sec:2a}

The effective field theory description for this system is valid for
momenta $p\alt\Lambda_S$ and consists solely of contact interactions
in a derivative expansion between the nucleons.  
The fundamental $NN$ vertex at tree-level is%
\footnote{The $\wh C$
  couplings used throughout are a factor of $M/4\pi$ times the conventionally 
  normalized $C$'s found in the literature.}
\beq
\CA_{\rm tree}^{\rm EFT} = -{4\pi\over M}\; \wh C(p^2) = -{4\pi\over
M}\(\wh C_0 + \wh C_2 p^2 + \wh C_4 p^4 +\ldots\)\, .
\eeq
By using dimensional regularization and the PDS subtraction scheme,
it is possible to compute the exact
scattering amplitude for the EFT in terms of $\wh C$,
resulting in \cite{KSW}
\beq
\CA^{\rm EFT} = -\frac{4\pi/M}{{\wh C^{-1}(p^2;\mu)}+ \mu+i p}\ ,
\eqn{ampseff}
\eeq
where $\mu$ is the renormalization scale.
Comparing with \eq{ampdef} we see that the EFT formally agrees
with the exact result, provided that the
$\wh C$ couplings are chosen to
satisfy
\beq
\frac{1}{\wh  C(p^2;\mu)}+\mu =-   p\cot\wt \delta \ .
\eqn{cval}
\eeq
Expanding both sides of \eq{cval} in powers of $p^2$ and employing
the effective range expansion \eq{ere}, solutions for
the running couplings $\wh C_{2n}(\mu)$ can be obtained.
The first few are
\beq
\wh C_0(\mu) &&= -{1\over {\mu-1/\wt a}}\ ,
\nonumber\\ 
\wh C_2(\mu)&&= \frac12 \wt r_0 \;[\wh C_0(\mu)]^2 \ ,
\eqn{rgsol}
\\
\wh C_4(\mu)&&={[\wh C_2(\mu)]^2\over \wh C_0(\mu)} 
+ \wt v_2 \;[\wh C_0(\mu)]^2 \ .
\nonumber
\eeq
Two important features of these solutions are:
\begin{itemize}
\item The EFT exhibits a nontrivial unstable infrared fixed point 
at $\mu\,\wh C_0 = -1$ and $\mu^{2n+1} \wh C_{2n} =0$. This fixed point 
corresponds to a system with infinite scattering length, for which
scattering near threshold due to $\wh C_0$ looks scale invariant, in
spite of the fact that the $\wh C_0$ interaction is irrelevant by
naive power counting~\cite{birse1}.
\item Only for $\mu\sim \Lambda_S$ are the $\wh C_{2n}$ parameters of the
  size one would expect from naive power counting \cite{Weinberg1},
  assuming that the scattering parameters $\{\frac12\wt r_0, \wt
  v_n\}$ of the short range interactions are of natural size; $\wh
  C_0$ at this scale is fine tuned so that the system passes near the
  nontrivial infrared fixed point.
\end{itemize}
Although in the present example the PDS scheme allows for a formal
summation of all the diagrams of the EFT, more realistic interactions require
a systematic expansion which approximates the exact EFT result
\eq{ampseff}.  We now explore two different expansion schemes.

\subsection{The Weinberg Expansion of the Effective Theory}
\label{sec:2b}

As discussed in \S\ref{sec:1}, the Weinberg approach
\cite{Weinberg1,KoMany,Parka,steelea} 
consists of expanding the kernel of the Lippmann-Schwinger
equation, and then iterating it to all orders.  
The coefficients of the dimensionful contact interactions are taken to 
scale with $\Lambda_S$ according to their naive dimension, namely
\beq
{\wh C_{2n}}p^{2n}\sim (1/\Lambda_S) (p/\Lambda_S)^{2n}\ .
\eqn{wexp}
\eeq
The kernel is then expanded in powers of $p/\Lambda_S$, where $p$ is the
external momentum, so that counting powers of $p$ gives a way
of organizing corrections to the potential.
At $N^{th}$ order in the expansion, the kernel is given by
$V^{\rm EFT}_{N} \equiv \wh C^{[N]}(p^2;\mu) 
=\sum_{n=0}^{N}\wh C_{2n}p^{2n}$.

Although traditionally the Weinberg approach is analyzed with a
momentum cutoff, we can equally well use dimensional regularization
with the PDS subtraction scheme \cite{KSW}.
At $N^{th}$ order in the expansion,
the amplitude is  given by \eq{ampseff}, with the exact interaction $\wh
C(p^2;\mu)$ replaced by  $\wh C^{[N]}(p^2;\mu)$.  The $N$
couplings $\wh C_{2n}$ can be fixed by requiring that the theory correctly
reproduce the first $N$ terms in the effective range expansion.
The predictions for the phase shift $\wt \delta$ at leading order (LO),
next-to-leading order (NLO), and next-to-next-to-leading order (NNLO) 
are then given by
\beq
(p\cot\wt\delta)_{\rm LO} &=& -{1\over \wt a}\ ;
\nonumber\\
(p\cot\wt\delta)_{\rm NLO} &=& (p\cot\wt\delta)_{\rm LO} 
+ { \frac12 \wt r_0 p^2\over
1+\,\frac12 \eta\, \wt r_0 p^2}\ ;\eqn{wres}
\\
(p\cot\wt\delta)_{\rm NNLO} &=& (p\cot\wt\delta)_{\rm NLO}
+ \frac{p^4
\( \wt v_2  + 
      \,\frac14 \eta\,  \wt r_0^2  \)}{ 
     \( 1 + \,\frac12 \eta\,\wt r_0 p^2 \) \,
     \( 1 + \,\frac12 \eta\, \wt r_0 p^2 + 
       \,\eta\, p^4 \[  \wt v_2 + 
          \,\frac14\eta\,\wt r_0^2\] \)} \ ,
\nonumber
\eeq
where $\eta\,\equiv (1/\wt a-\mu)^{-1}$. 
The explicit $\mu$ dependence in the physical quantities is a result
of neglecting counterterms 
necessary for renormalization at any finite order in the expansion.

For small $\mu$,
these expressions fail above $p\sim (\wt a\wt r_0)^{-1/2}$
which is much below the expected EFT breakdown of
$p\sim\Lambda_S$ (e.g., see the  $\mu=0$ treatment in
Ref.~\cite{KSWa}, or the equivalent momentum subtraction used in
Ref.~\cite{Weinberg1}).  This failure is no surprise, since for small $\mu$ 
the $\wh C_{2n}$ couplings given in \eq{rgsol} do not obey the naive
scaling assumed in \eq{wexp}, which was the starting point for Weinberg's
expansion. Evidently, the assumed power counting is only
obtained for
$\mu\gtap \Lambda_S$, in which case we see that the result in \eq{wres} 
coincides with the effective range expansion up to higher order terms%
\footnote{It is curious that in the limit $\mu\to \infty$, the
Weinberg expansion to $N^{th}$ order exactly reproduces the first $N$
terms of the effective range expansion for $p\cot\wt \delta$.}
in $p^2/\Lambda_S^2$.

\subsection{The KSW Expansion of the Effective Theory}
\label{sec:2c}

The KSW expansion starts from the assumption $\wt a\gg 1/\Lambda_S$ 
and that the rest of the effective range parameters $\{\frac12\wt r_0,\wt
v_n\}$ scale with appropriate
powers of $\Lambda_S$, according to their dimension.
In this case, the amplitude \eq{ampdef} may be expanded as
\beq
\CA = -{4\pi/M\over (1/\wt a + i p)}\[ 1 + {\wt r_0/2 \over
(1/\wt a + ip)}p^2 + {(\wt r_0/2)^2\over (1/\wt a + ip)^2} p^4 + {\wt
v_2 \over (1/\wt a + ip) } p^4 
+\ldots\]\ ,
\eqn{kswexp}
\eeq
provided one is not near the kinematic point corresponding to a pole
in $\CA$. 
This expansion can be realized in
the EFT by applying the following power counting rules 
for $s$-wave interactions \cite{KSW}:
\begin{enumerate}
\item Define a measure of small momenta called $Q$, and take $p$, 
    $\mu$,  and $1/\wt a$ to be $\order(Q)$.
\item The $\wh C_{2n}(\mu)$ couplings, as given in \eq{rgsol}, scale as
  $\order(Q^{-(n+1)})$.
\item Loop momenta $\bfq$ are $\order(Q)$, while loop energies $q_0$ are
  $\order(Q^2)$.  As a result, loop integrals (including two-nucleon
  propagators) scale as $\order(Q)$.
\item Derivative interactions contribute a power of $Q$ for each
  $\nabla$, and $Q^2$ for each $\partial_t$.
\end{enumerate}
Summing up all Feynman diagrams to a given order in $Q$ and
making use of \eq{rgsol} reproduces the expansion \eq{kswexp}.
Since this is a consistent expansion of a physical quantity, results are 
$\mu$-independent at each order.
The most peculiar feature of the expansion is the
scaling of $\wt a$ as $\order(Q^{-1})$.  
This allows for an expansion
in powers of $p/\Lambda_S$ while keeping powers of $p\wt a$
to all orders, but it will cause subtleties when the scale $m_\pi$ is 
introduced.

\section{Adding Toy Pions}
\label{sec:3}

\begin{figure}[t]
\centerline{{\epsfxsize=3.0in \epsfbox{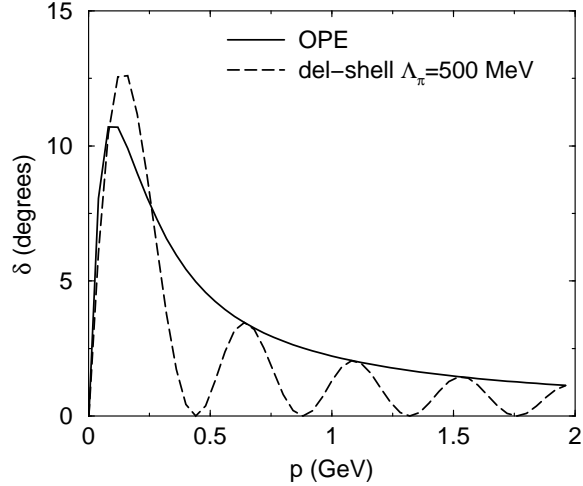}} }
\noindent
\caption{\label{fig:lambda} By taking $\Lambda_\pi=500$~MeV, the
envelope of the OPE and delta-shell potential coincide.}
\vskip .2in
\end{figure}

We now consider the more interesting case of an interaction with both
long- and short-range structure. 
The short-range interaction will remain
unspecified for generality, 
giving the phase shift $\wt \delta$  when acting alone.  
For the long-distance part
of the potential $V_L$, we choose the specific form of a delta-shell
at a radius given by the pion Compton wavelength and a
strength depending on $m_\pi$ in a manner consistent with chiral
symmetry 
\beq
V_L(r) = - {g_\pi m_\pi\over M}\; \delta\!\(r-\frac1{m_\pi}\) 
\ .
\eqn{vldef}
\eeq 
This caricature of the real one-pion exchange (OPE) potential 
allows us to analytically explore issues involving EFT expansions.
The  Yukawa part of the $\si$ channel OPE
interaction in momentum space is 
\beq
\tilde  V_{\rm OPE} = - {4\pi\over M\Lnn} 
\;{m_\pi^2\over q^2 + m_\pi^2}\ ,
\qquad 
\Lnn\equiv {16\pi f_\pi^2\over g_A^2 M}
\simeq 300~{\rm MeV},
\eqn{opep}
\eeq
where $m_\pi=140$~MeV is the pion mass, $f_\pi=93$ MeV is the pion
decay constant, and $g_A=1.29$ is the axial coupling.  
The Fourier transform of $V_L$ is
\beq
\tilde V_L = -{4\pi g_\pi\over M m_\pi}\; {\sin q/m_\pi\over
q/m_\pi}\, .
\eqn{momvldef}
\eeq
Comparison of the two expressions suggests that for $V_L$ to have
comparable effects to $V_{\rm OPE}$, we should take
$g_\pi=m_\pi/\Lambda_\pi$ with $\Lambda_\pi \sim \Lnn$.
In fact, the properties of $V_L$ and the OPE Yukawa potential are
most analogous if we take 
\beq
\Lambda_\pi = 500~{\rm MeV}\ .
\eeq
By analogous, we mean that in both cases, the strength of the
interaction is approximately 30\%  
of the critical value needed to give 
a bound state (which occurs at $g_\pi=1$ in the toy model), 
and the strength of the phase shifts resulting from the 
two potentials are roughly comparable as shown in \fig{lambda}.

With $\Lambda_S > m_\pi$, it is straightforward to solve the
Schr\"odinger equation
\beq
\[-{\nabla^2\over M} + V_S + V_L\] \psi = E \psi
\eqn{seq}
\eeq
and determine the exact $s$-wave Feynman scattering amplitude $\CA$ to be
\beq
\CA &&= \CA_L + \frac{4\pi/M}{ f(\xi)^2 \(p
\cot\wt\delta - i p\) - g_\pi m_\pi e^{2i\xi}f(\xi)}\ ,
\eqn{ampfull}
\eeq
with $\xi=p/m_\pi$ and 
\beq
f(\xi) &&\equiv 1-g_\pi {e^{i\xi} \sin\xi\over \xi}\ .
\eqn{fdef}
\eeq
As in the previous section, $\wt \delta$ is the phase shift due to
$V_S$ alone. The quantity $\CA_L$ is the amplitude one finds for $V_L$
alone, when the short-range potential $V_S$ is turned off,
\beq
\CA_L = {4\pi g_\pi\over M m_\pi} \frac{\sin^2\!\xi}{f(\xi) \;\xi^2}
\, .
\eqn{amppi}
\eeq
The exact phase shift $\delta$ can be found to be
\beq
\cot\delta = \;\frac{(\xi-g_\pi \sin \xi\cos \xi)\cot \wt\delta - g_\pi
\cos^2 \xi}{g_\pi \sin^2\!\xi \;\cot \wt\delta + \xi +
g_\pi\sin\xi\cos\xi} \ . 
\eqn{exactps}
\eeq
{}From this, one can determine the exact effective range parameters
defined by
\beq
p\cot\delta=-\frac1{a} + \frac12 r_0 p^2 + 
\sum_{n=2}^\infty v_n p^{2n} + \ldots
\eeq
in terms of the short-distance effective range parameters \eq{ere}.
For example, with $g_\pi\equiv m_\pi/\Lambda_\pi$,
\beq
a&=&{1-\wt a (\Lambda_\pi + m_\pi)\over m_\pi(1-\wt a m_\pi) -
\Lambda_\pi}\, ,
\\  \nonumber \\
r_0&=&\frac{2m_\pi \( 1 - \wt a m_\pi \)^2 + 
     2\Lambda_\pi \( 1 - 3\wt a m_\pi\) \(1-\wt a m_\pi\) +
     3 {\wt a }^2 \Lambda_\pi^2 m_\pi^2 \wt r_0 }
     {3 m_\pi^2 \[ 1 - \wt a \( \Lambda_\pi + m_\pi \)  \] ^2}\ .
\eqn{erexp}
\eeq
We wish to reproduce the full amplitude $\CA$ in \eq{ampfull} 
in a systematic 
expansion of an EFT, replacing the potential $V_S$
with contact interactions.  
We first show that the Feynman
amplitude can be exactly reproduced by the effective theory, as was
the case when we considered $V_S$ alone.  
Subsequently, we address the issue
of how to compute these amplitudes in perturbation theory, which is of
practical interest for realistic $NN$ scattering, where exact 
analytical solutions to the EFT are impossible to obtain. 
 
\subsection{The Exact Solution of the Effective Theory with Toy Pions}
\label{sec:3a}

Taking an EFT with contact interactions $\wh C(p^2;\mu)$ 
and the explicit long-range potential $V_L$ given in \eq{vldef},
it is again possible to sum up all diagrams using 
dimensional regularization
and the PDS subtraction scheme (see Appendix~\ref{sec:6}) resulting in
\beq
 \CA^{\rm EFT}=\CA_L  +  {4\pi/M\over  -f(\xi)^2  \[ 
 \wh C^{-1}(p^2;\mu) +\mu+i p\] - g_\pi m_\pi  e^{2i\xi} f(\xi)}\ ,
\eqn{ampfulleft}
\eeq
with $f(\xi)$ and $\CA_L$ as in \eqsii{fdef}{amppi}.
A comparison between the above expression and the exact amplitude
\eq{ampfull} shows that the EFT reproduces the full theory provided
the $\wh C$ couplings are chosen to satisfy
\beq
\frac{1}{\wh  C(p^2;\mu)}+\mu =-   p\cot\wt \delta \
.
\eqn{cvaleft}
\eeq
This is exactly the same result that we derived earlier in the
theory without long-range interactions, \eq{cval}.  
Thus, the $\wh C_{2n}$ interactions truly represent the short-distance
physics alone, and are independent of $V_L$. (This need not in
  general be true for the more realistic case of nonzero overlap 
in space between $V_S$ and $V_L$.)

\subsection{The Weinberg Expansion with Toy Pions}
\label{sec:3b}

In Weinberg's power counting scheme, the scale $\Lambda_\pi$ is
considered a short-distance scale and so 
insertions of $V_L$ are of the same order as insertions of $\wh C_0$,
as seen from \eq{momvldef}.  
Thus to $N^{th}$ order in the expansion, the effective potential is
given by%
\footnote{In general, these operators
have to be specified more precisely, for example $\wh C_2 p^2$ must be
replaced by $\frac12 \wh C_2 \{\hat p^2,\delta^3(\hat r)\}$ and so on.}  
\beq
V^{\rm EFT}_{N} = V_L + \sum_{n=0}^{N} \wh C_{2n} p^{2n}\, .
\eqn{trunc}
\eeq
The
$N^{th}$ order expression for the Feynman amplitude is just given by
the exact EFT solution \eq{ampfulleft} with $\wh C(p^2;\mu)$ replaced by the
truncated series of $\wh C_{2n}p^{2n}$ in \eq{trunc}.  The discussion now is
entirely analogous to that of \S\ref{sec:2b}.  
Again the amplitude is not renormalized and therefore is $\mu$-dependent.
However, as long as $\mu\gtap \Lambda_S$ and the $N$ couplings 
$\wh C_{2n}$  are chosen to reproduce the effective range expansion (up to
higher order terms), the exact expression for the amplitude
\eq{ampfull} is obtained with $p\cot\wt \delta$ replaced by the first 
$N$ terms in the short-distance effective range expansion. 
Note that Weinberg's expansion is functionally equivalent to the
modified effective range expansion \cite{modER1} as long as the
overlap between $V_L$ and $V_S$ is properly accounted for, as 
discussed in Ref.~\cite{steeleb}.

\subsection{The KSW Expansion with Toy Pions}
\label{sec:3c}

\subsubsection{The Naive KSW Expansion and Its Failings Near the Pole}
\label{sec:3ci}

Once pions are included in the EFT, the KSW expansion treats $m_\pi$ as
order $Q$ as well, generalizing the rules of \S\ref{sec:2c}.  
This is consistent with chiral perturbation theory, which treats $p$
and $m_\pi$ to be of the same order and small.  
It follows that $V_L$ in \eq{momvldef} is
$\order(Q^0)$, which is subleading compared to $\wh C_0$, which is
$\order(Q^{-1})$. 
Therefore, pions
enter the amplitude perturbatively \cite{KSW}, 
unlike in the Weinberg approach.
It is advantageous to treat the OPE potential perturbatively, since
perturbative long-range interactions are analytically
tractable, at least up to NNLO.
While this is not a major  benefit in two-nucleon systems, 
it is likely to be very valuable in many-body problems due to
computational simplifications.  

However, there is a subtlety associated with implementing this method, 
having to do with the large scattering length which arises from the 
delicate interplay between short-range physics and pion exchange.  
Making use of the exact
solution for $\wh C_0$ from \eq{rgsol}, one finds that the
leading contribution to the amplitude at $\order(Q^{-1})$ is 
\beq
\CA^{\rm EFT}_{-1} = - {4\pi/M\over 1/\wt a + i p}\ .
\eqn{leading}
\eeq
This has a pole at $p_\star=i/\wt a$, in contrast to the exact
amplitude which has a pole at $p_\star \simeq i/a$. Even if $1/a$ and
$1/\wt a$ differ by an amount small compared to $m_\pi$, since the
threshold cross section is $4\pi a^2$, 
 this expansion will fail to accurately describe  low-momentum
 scattering if either  
$a$ or $\wt a$ are particularly large. 
For example, in the extreme
case where the scattering length 
$a$ is infinite, then the true amplitude at threshold is
also infinite.  
However, $\wt a$ would be finite in that case, and so \eq{leading}
would require the subleading perturbative corrections to the cross section to
be infinite in 
order to produce the correct result!

The situation is similar to the problem of calculating $e^+
e^-$ scattering amplitudes at the $Z^0$ pole in perturbation theory.
Even though the true 
value for the $Z^0$ mass and the tree level value only differ by
an $\order(\alpha)$ contribution,
practical calculations are done by perturbing around the exact
$Z^0$ mass and introducing a mass shift at each order in perturbation
theory to cancel the subsequent radiative corrections. 

Similarly, 
in the present example, a small difference between $1/\wt a$
and $-i p_\star$  does not ensure an accurate description of
scattering for  $p\sim \vert 1/a\vert $.  
For large and positive $a$, 
where a shallow bound state exists and the pole in $\CA$ is
on the physical sheet (such as the deuteron pole in $\siii$  $NN$ scattering), an
expansion around the true location of the pole is necessary.
For large and negative $a$, where the pole is kinematically inaccessible
(as in the $\si$ channel), it suffices to
reproduce the exact scattering length at leading order.
This is the procedure of matching at LO already carried out in the
literature \cite{KSW}, but it needed to be understood from the perspective
of the short-distance scales given in \eq{rgsol}, because extending the
procedure to NLO differs from the conventional approach. The following 
discussion has much in common with that of Mehen and Stewart
\cite{MS1} and Rupak and Shoresh \cite{RS}.

\subsubsection{The Pole Expansion}
\label{sec:3cii}

The resolution, as in the example of the $Z^0$ pole, is to reorder the 
expansion, making use of the fact that the difference between $1/\wt
a$ and either $1/a$ or $-i p_\star$ is $\order(Q^2)$.  For the case
relevant to $NN$ scattering in the $\si$ channel, we write
\beq
{1\over \wt a} = {1\over a} + \sum_{k=2}^\infty \alpha_k\ ,
\eqn{scats}
\eeq
where the $\alpha_k$ are $\order(Q^k)$ functions of  $a$,
$m_\pi$, $\Lambda_\pi$ and $\mu$.
For example, the long-range potential
we are considering gives
\beq
{1\over \wt a} =  {\frac{\Lambda_\pi + m_\pi - a\,{m_\pi^2}}
   {1 + a\,\Lambda_\pi - a\,m_\pi}}
\qquad
\Longrightarrow
\qquad
\alpha_n = - \Lambda_\pi \({a m_\pi -1\over a \Lambda_\pi}\)^n\ .
\eqn{alphadef}
\eeq
Substituting the expression for the short-distance scattering length
\eq{scats} into $\wh C_0$ from \eq{rgsol} gives an expansion in $Q$:
\beq
\wh C_0(\mu)
\equiv \sum_{k=-1}^\infty \wh C_0^{[k]}(\mu)\ , 
\eqn{c0exp}
\eeq
where $\wh C_0^{[k]}$ is of order $Q^k$.  The first three contributions are
\beq
\wh C_0^{[-1]} =-{1\over \mu-1/ a}\ ,\qquad 
\wh  C_0^{[0]}= -\alpha_2 
\(\wh C_0^{[-1]}\)^2\, , \qquad
\wh C_0^{[1]} = \alpha_2^{\ 2} \left( \wh\Cs{0}{-1} \right)^3 - \alpha_3
\left( \wh\Cs{0}{-1} \right)^2\, .
\eqn{c0exp1}
\eeq
This expansion ensures that the EFT yields the exact
scattering length $a$ at each
order in the KSW expansion, with $\wh C_0^{[k]}$ corrections to cancel
the perturbative contributions of the pions arising at each order.  
Note that $\wh C_0$ as defined in
\eq{c0exp} remains $m_\pi$-independent, even though the $\wh C_0^{[k]}$ 
terms each have a rather complicated $m_\pi$ dependence. 

Once $\wh C_0$ is expanded, the exact solutions for
the $\wh C$ couplings in \eq{rgsol} imply that
each $\wh C_{2n}$ coefficient must be similarly expanded. 
For instance,
\beq
\begin{array}{rclrcl}
\wh C_2^{[-2]} &=& \frac12 \wt r_0 \(\wh C_0^{[-1]}\)^2\ ,&\quad 
\wh C_2^{[-1]} &=& \wt r_0 \(\wh C_0^{[-1]}\wh C_0^{[0]}\)\ ,
\quad \ldots\, ,
\\  \\
\wh C_4^{[-3]} &=& \displaystyle
\frac{\(\wh C_2^{[-2]}\)^2}{\wh C_0^{[-1]}}\ ,&\quad 
\wh C_4^{[-2]} &=& \displaystyle
2 \frac{\wh C_2^{[-1]}\wh C_2^{[-2]}}{\wh C_0^{[-1]}} +
\alpha_2 \(\wh C_2^{[-2]}\)^2
+ \wt v_2 \(\wh C_0^{[-1]}\)^2\ ,
\quad \ldots\, . \\
\end{array}
\eqn{c2exp}
\eeq
That the leading behavior of each $\wh C_{2n}$ is determined by
lower dimensional operators is further evidence of the EFT being
tuned to lie near a nontrivial fixed point. 

Note that in the KSW expansion, the parameter $\wt r_0$ first enters 
at $\order(Q^0)$ through an insertion of 
$\wh C_2^{[-2]}p^2$, while the 
next new parameter $\wt v_2$ does not enter until $\order(Q^2)$
through an insertion of $\wh C_4^{[-2]} p^4$. 
In general, the short-distance effective range parameters \eq{ere}
enter the expansion of $\CA$ through the operators
$\wh C_{2n}^{[-2]}p^{2n} \sim \order(Q^{2n-2})$ for each $n\ge 1$.  
Thus in the KSW expansion, one new parameter is encountered for
every {\it two} powers of $Q$. 
The NNLO amplitude $\CA_1$, for example, involves graphs not
included in $\CA_0$;
nevertheless, it is completely parameterized by the same two numbers $a$
and $\wt r_0$.  
The fact that new parameters appear in the expansion of $\CA$ with only
even powers of $Q$ occurs because the effective range
expansion of $p\cot\wt\delta$ is in even powers of momentum.  
A valid fitting procedure
for the EFT should reflect this behavior, with one free 
parameter at LO to fix the scattering length $a$, one free parameter
at NLO to fix $\wt r_0$, and subsequent free parameters
appearing at every other order
$\order(Q^{2n-2})$ to fix each $\wt v_{n\ge2}$.

\subsubsection{A General Algorithm for Fixing Coupling Constants}
\label{sec:3ciii}

The Weinberg expansion has a relatively simple algorithm for fixing the
low-energy constants of the EFT: Taylor expand the EFT result for
$p\cot\delta$ and match to the effective range expansion at each order
(making sure the long-distance effects are properly taken into account
\cite{steeleb}).
The procedure to fix the low-energy constants in the KSW expansion is
more obscure. When the nucleon interactions happen to be fine-tuned, high
order perturbative pion contributions can in principle have relatively 
large effects on low-energy observables due to cancellations at lower orders 
in the expansion.\footnote{For the toy models we are considering, the
  Weinberg expansion does not suffer from this problem, since the
  long-range interactions are included completely at lowest order.
  However, for real $NN$ scattering, the Weinberg method also
 involves a chiral expansion of the long-range interactions beyond
 OPE, and so in 
  principle subleading contributions could similarly exhibit large effects on
  low-energy scattering, complicating the determination of the contact
interactions $\wh C_{2n}$.}
Our simple long-distance potential $V_L$ allows for an analytical
determination
of all the $\wh C_{2n}^{[k]}$ couplings (as we will show in an example in
\S\ref{sec:4}).  
However, for realistic
situations one needs an algorithm for fixing these couplings. 
The above discussion makes it possible to surmise a general prescription:%
\footnote{Here we describe the procedure for a system without a bound
  state, such as $NN$ scattering in the $\si$ channel. 
  Systems with a bound state, where one needs to fix the pole
  $ip_\star$ at LO instead of the scattering length $a$, are considered 
  elsewhere \cite{Kaplan:1998sz,RS}. Our procedure can easily be
extended to these cases.}

\begin{enumerate}
\item Expand each $\wh C_{2n}$ coupling as $$\wh C_{2n} =
    \sum_{k=-(n+1)}^\infty \wh C_{2n}^{[k]}\ ,$$ where 
    $\wh C_{2n}^{[k]}$ is
    $\order(Q^k)$ in the expansion, and compute the amplitude to the desired
    order.
\item Use the threshold amplitude to fix 
  $\wh C_0^{[-1]}$ so that the LO result reproduces the experimental
  scattering length $a$, and fix the higher order 
  $\wh C_0^{[k]}$ by requiring the scattering length
  be unchanged at each higher order.
\item Determine the renormalization group equations in the PDS scheme
for the $\wh C_{2n}$ couplings to the order one is working.  The beta function
  for $n\ge 1$ is always of the form 
\beq
\mu \frac{d \wh C_{2n} }{d\mu} = 2 \mu\wh C_0\wh C_{2n}+
\mu \frac{d F_{2n}}{d\mu}
\eqn{f2n}
\eeq
where $F_{2n}$ only depends on couplings $\wh C_{2m}$ with $m<n$ and
possibly the long-distance physics.%
\footnote{The toy pions \eq{vldef} are well behaved at the origin and
therefore never
contribute to the $F_{2n}$'s.  Actual pion exchange has a finite
overlap with the EFT contact interactions, so $\log \mu$ terms could
occur.  However, at least to NNLO in the $Q$-expansion, only $F_{0}$ has any
such contribution.}
Note that for $n>1$, the $F_{2n}$ are actually the leading
contribution in the $Q$-expansion of \eq{f2n}. 
Solving this equation leads to
\beq
\wh C_{2n} = \wt v_{n} \wh C_0^{\ 2} + F_{2n}\ ,
\eeq
where $\wt v_n$ is an undetermined constant of integration treated as
$\order(Q^0)$. 
This solution should then be expanded in $Q$ to produce each of
the $\wh C_{2n}^{[k]}$'s to be used in the calculated amplitude $\CA$.
\item Within the $Q$-expansion, the phase shift is computed
  consistently from this amplitude $\CA$
  in terms of the parameters $\{\frac12\wt r_0,\wt v_n\}$, which are the
  short-distance effective range parameters.
  These parameters are then determined by finding the
  best fit to the data with an appropriate weighting in momentum
  (as discussed in \S\ref{sec:4}).
\end{enumerate}
In this manner, contact is made between the plethora of terms in the
KSW expansion, and the effective range parameters 
\{$\frac12 \wt r_0$, $\wt v_n$\} 
characterizing the short distance physics. At a given order,
the KSW and Weinberg expansions have the same number of free
parameters to fit.
This general algorithm will correctly reproduce the KSW expansion
in the presence of pions as explicitly shown in \S\ref{sec:4}.

Note that additional contact interactions with quark mass insertions
are possible \cite{KSW}, such as $D_2 m_\pi^2$ at NLO and
$D_{22}m_\pi^2p^2$ at NNLO. However, to distinguish
between these operators and the  $C^{[k]}_{2n}$
interactions would require additional information beyond $NN$
scattering, such as $\pi d$ scattering.  Therefore, in the present
discussion we subsume them 
into our definitions of the $C^{[k]}_{2n}$'s.

\section{Examples}
\label{sec:4}
We now illustrate the above procedure for fixing coupling constants in
the KSW expansion scheme by three examples.
Choosing an explicit form for the short-distance potential allows us
to check that this method is actually working.
In all cases, we will tune the couplings in the 
model to give $a=-23$~fm similar to the $\si$ channel of $NN$
scattering.

\subsection{The Toy Pion with a Short-range Delta-shell Interaction}
\label{sec:4a}

Our first example will utilize the toy pions of \eq{vldef}.
We choose the short-distance physics to also be modeled by a
delta-shell so an analytic solution is possible
\beq 
V=V_S + V_L\ ,\qquad 
V_S(r) = - g_\rho {m_\rho\over M} \delta\left(r-{1\over
m_\rho}\right)\ ,\qquad
V_L(r) = - {m_\pi^2\over M\Lambda_\pi} \delta\left(r-{1\over
m_\pi}\right)\ ,
\eeq
with $\Lambda_\pi=500\MeV$.
The exact phase shift is then just given by \eq{exactps} with
\beq
p\cot\wt\delta = \frac{\xi_\rho-g_\rho \sin\xi_\rho\, \cos\xi_\rho}
{g_\rho \sin^2\xi_\rho}\, ,
\eqn{shortyuk}
\eeq
and $\xi_\rho=p/m_\rho$.  Taking $m_\rho=770$~MeV, 
we choose $g_\rho=0.915$ to give the  scattering length $a=-23$~fm.

The EFT amplitude $\CA$ up to NNLO is quoted in Appendix~\ref{sec:7}.
The contact interactions $\wh C_{2n}^{[k]}$ were worked out in general in 
\eqsii{c0exp1}{c2exp}, so all that is left to do is implement the
procedure outlined in \S\ref{sec:3ciii} to fix these constants.
The EFT calculation at LO only depends on $\wh C_0^{[-1]}$, which 
as seen from \eq{c0exp1} is fixed by the experimental scattering length.

At NLO, pions begin to contribute, and the perturbative correction
$\wh C_0^{[0]}$ is chosen to ensure
the scattering length does not shift.  
This specifies $\alpha_2$, which coincides with the analytic
result given in \eq{alphadef}. 
The contact interaction $\wh C_2^{[-2]}$ also contributes at this order
and depends on $\wt r_0$.
We use its exact value, which can be calculated from \eq{shortyuk}
\beq
\wt r_0 = \frac{2(1+g_\rho)}{3 g_\rho m_\rho}  
=0.36 \mbox{\ fm}\, ,
\eqn{r0ds}
\eeq
to test our fit.  Afterwards, we will discuss how to find this value when the
short-distance physics is not  known {\it a priori}.

At NNLO, choosing $\wh C_0^{[1]}$ so the scattering length does not shift
produces $\alpha_3$ as given by \eq{alphadef}.
The other new contact interactions $\wh C_2^{[-1]}$ 
and $\wh C_4^{[-3]}$ are
fully determined by already specified quantities \eq{c2exp}, so there
are no new constants to fix at this order as already discussed.

\begin{figure}
\begin{center}
\leavevmode
\hbox{
\hspace{-.5cm}
\epsfxsize=3.4in
\epsffile{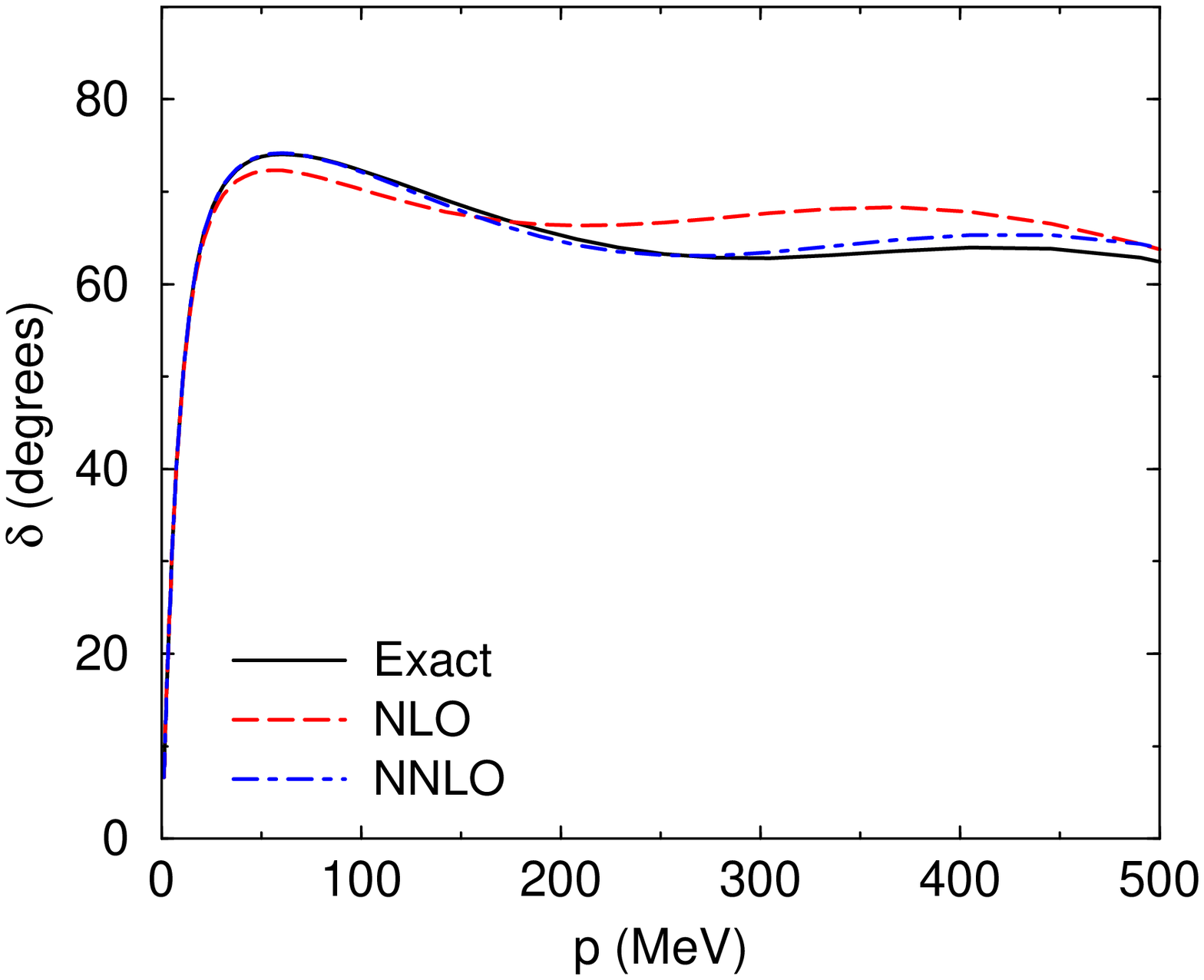}
\hspace{-.25cm}
\epsfxsize=3.4in
\epsffile{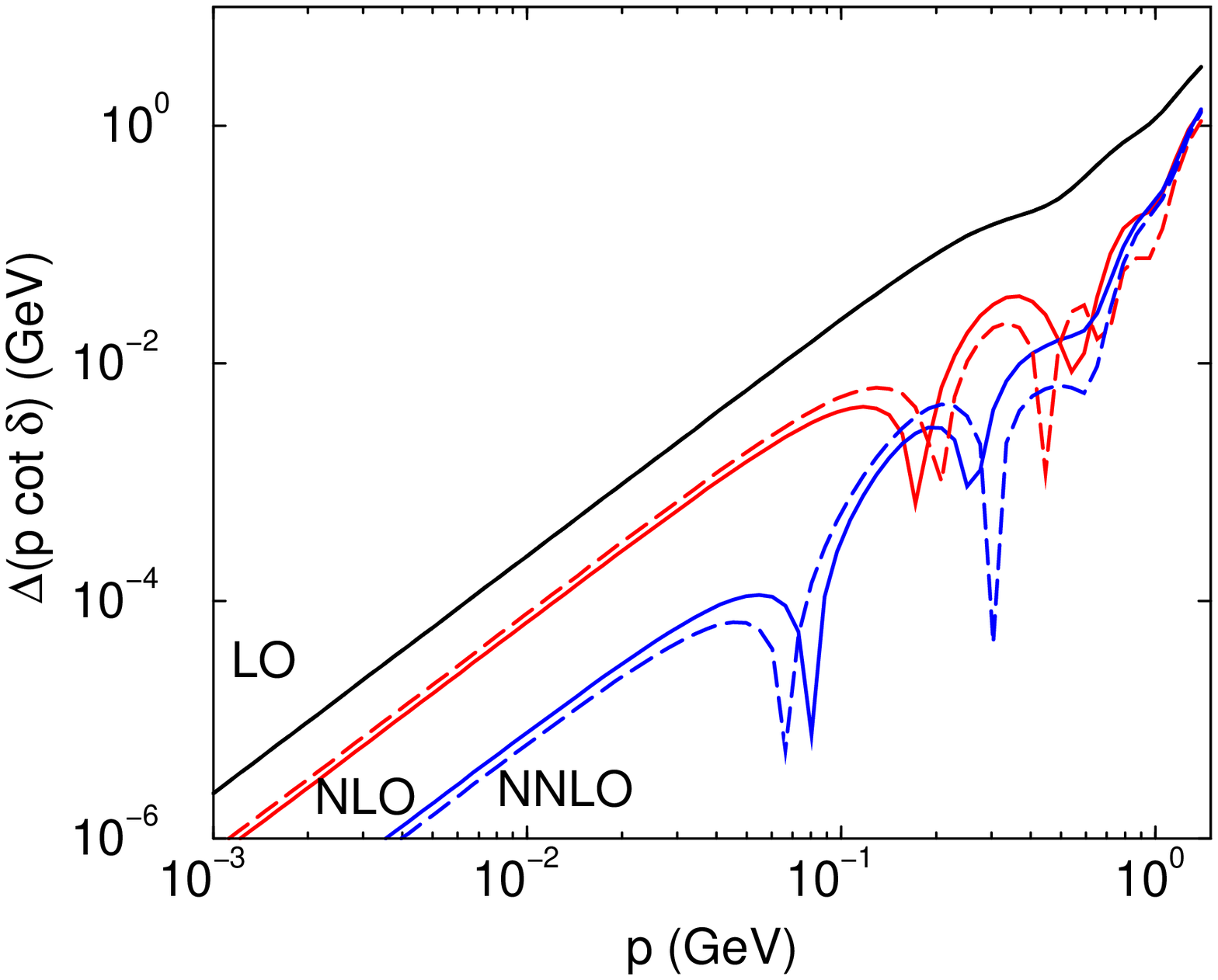}
}
\end{center}
\caption{\label{fig:2ds}The algorithm of \S\ref{sec:3ciii} 
for the KSW expansion
applied to the two delta-shell model.  The left plot shows that as
more orders are added 
to the EFT calculation, the agreement with the exact 
phase shift improves. The right plot shows the corresponding
errors in the observable $p\cot\delta$.}
\end{figure}

The results are shown in Fig.~\ref{fig:2ds}. 
On the left, the phase shift as a function of momentum for
the exact result (solid line) is compared to the EFT calculation at
NLO (dashed) and NNLO (dot-dashed).
Note that the results improve as more orders are added, even when
there are no additional free parameters (as occurs at NNLO).
The right plot of Fig.~\ref{fig:2ds} shows the improvement in the
observable $p\cot\delta$ by plotting its error at each order (solid lines).

If the exact value for $\wt r_0$ were not known, as is the case in
real data, we could treat $\wt r_0$ as a free parameter of the
effective theory,  varying its
value to achieve the best global fit to 
data over a representative momentum range such as $[1/a,\Lambda_\pi]$.
Doing this at both NLO and NNLO produces the dashed lines in the
right plot of Fig.~\ref{fig:2ds}, coinciding quite well with the exact
result.
Another indication of this agreement can be found by comparing the fit values
for $\wt r_0$ with the exact value \eq{r0ds}.
The EFT with a global fit gives $\wt r_0=0.41$~fm at NLO 
and $\wt r_0=0.37$~fm at NNLO, 
showing convergence to the exact result $\wt r_0=0.36$~fm of
\eq{r0ds}, differing only by 
higher order contributions which the fit cannot resolve.

The point at which this fitting procedure breaks down can also be
determined.
The EFT should only work for momentum below a scale associated with
underlying physics not explicitly accounted for in the Lagrangian,
which is $m_\rho$ for this two delta-shell model.
Up to this point, we have computed the phase shift 
by expanding the expression~\cite{KSW}
\beq
\delta &=& 
\frac1{2i}\; \ln\!\left[ 1 + i \frac{Mp}{2\pi} \left( {\cal A}_{-1} +
{\cal A}_0 + \ldots \right) \right] 
\eqn{exp1}
\\
&\simeq& \frac1{2i}\; \ln\!\left( 1 + i \frac{Mp}{2\pi} {\cal A}_{-1}
\right) + \frac{Mp}{4\pi} \left( \frac{{\cal A}_0}{1+i \frac{Mp}{2\pi}
{\cal A}_{-1}} \right) + \ldots
\, ,
\eqn{exp2}
\eeq
but we could also have kept the full expression \eq{exp1}.
These two expressions are equivalent up to terms that are
higher order in the expansion.
Those extra terms are an estimate of the corrections to the
actual result and depend on the radius of convergence.
The breakdown of the KSW expansion can be identified as the
point at which the two expressions \eqsii{exp1}{exp2}
diverge from each other. 
Doing this exercise for the two delta-shell model using the results of
the fit to NLO and NNLO 
reveals the breakdown to be near
$m_\rho$ as expected \cite{steelea}.

The delta-shell model for the pion allows us to obtain
analytical expressions for the scattering amplitude and to check the
algorithm 
for determining the $\wh C_{2n}^{[k]}$ coefficients of the EFT as
described in \S\ref{sec:3ciii}.  
However, this toy model can be criticized as being too simple to provide
evidence that the KSW expansion will work for actual $NN$ scattering.  
We therefore consider models where pion exchange is
represented by the correct OPE potential for the $\si$ channel.

\subsection{The Two-Yukawa Model}
\label{sec:4b}

We now take a more sophisticated model 
with a long-distance potential given by the Yukawa part of 
the $\si$ OPE potential,  and a short-distance Yukawa potential
characterized by the rho mass
\beq
V=V_\pi + V_S\ ,\qquad V_\pi = - \alpha_\pi \frac{e^{-m_\pi r}}{r}\
,\qquad V_S=
- \alpha_\rho \frac{e^{-m_\rho r}}{r}\, .
\eqn{2yuk}
\eeq
The pion coupling is taken to be
$\alpha_\pi=g_A^2 m_\pi^2/(16 \pi f_\pi^2)\simeq0.075$,
as in the real world, and the rho coupling
$\alpha_\rho=1.05$ is tuned to give a large scattering length
$a=-23         $~fm, as observed in data for 
the $\si$ partial wave of $NN$ scattering.  Analyzing scattering from
the short-range 
$V_S$  potential alone then yields
\beq
\wt a = -1.17~{\rm fm}\ ,\qquad \wt r_0 = 0.765~{\rm fm}\ ,
\eqn{wt2yuk}
\eeq
and the effective range expansion for the phase shift $\wt \delta$
works well up to $p\simeq m_\rho/2$, as expected from
analyticity considerations.

\begin{figure}
\begin{center}
\leavevmode
\hbox{
\hspace{-.5cm}
\epsfxsize=3.4in
\epsffile{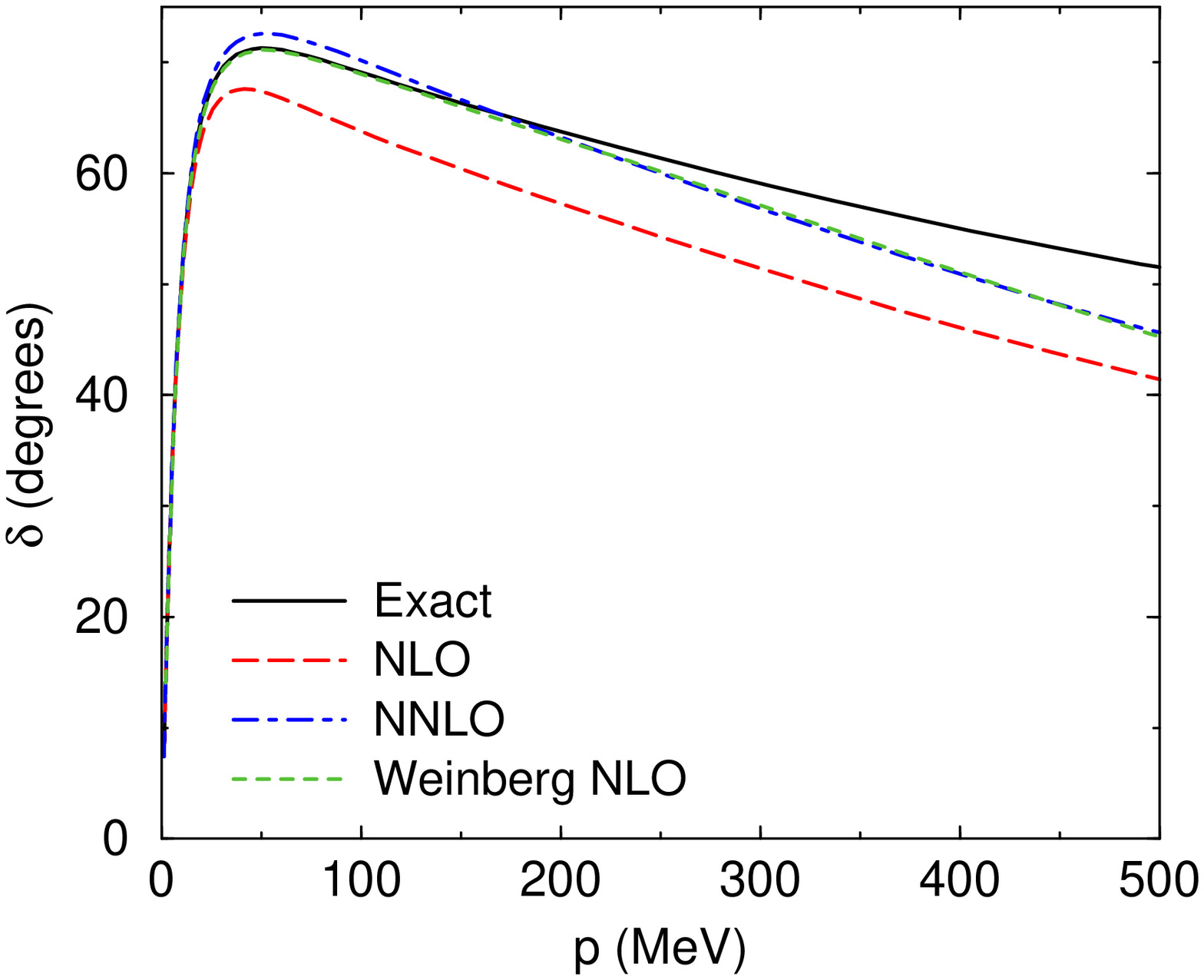}
\epsfxsize=3.4in
\epsffile{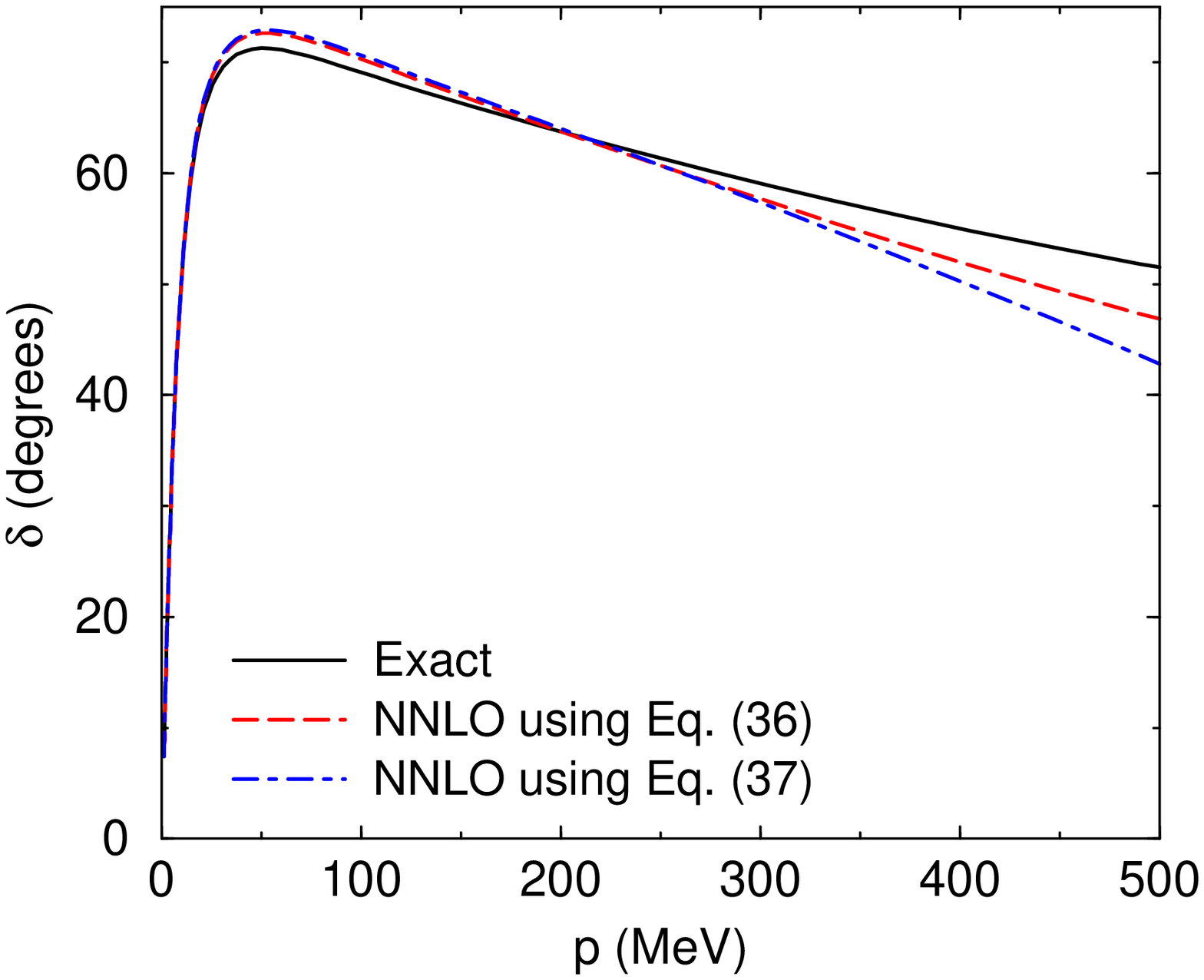}
}
\end{center}
\caption{\label{fig:2yuk}The algorithm of \S\ref{sec:3ciii} 
for the KSW expansion
applied to the two-Yukawa model.  The left plot compares NLO and NNLO
results from the EFT to the exact result.  The right plot shows 
the NNLO results using the different expansion schemes
\eqsii{exp1}{exp2} to determine the
breakdown scale, which appears to be around $400\MeV$.}
\end{figure}

The EFT amplitude with potential pions has been calculated up to
NNLO in Refs.~\cite{RS,MSpc}.  
We make use of that amplitude and fix the contact interactions according to
the algorithm described in \S\ref{sec:3ciii}.
This gives a LO amplitude that depends upon $a$, and 
NLO and NNLO amplitudes that also depend upon $\wt r_0$, the 
short-distance effective range parameter.  
The results obtained for the actual value  $\wt r_0=0.765$~fm
are shown in Fig.~\ref{fig:2yuk} for NLO and NNLO.  
The corresponding values for the full effective range are
$r_0=2.1$~fm and $r_0=1.3$~fm respectively, compared to the exact
value $1.6$~fm.

The result from using the Weinberg approach (short dashed line) is
also shown.  Since it is fit at low-momentum, the agreement there is
better than the KSW expansion.  Both expansion schemes coincide at
large momentum, which is in agreement with the idea that the Weinberg
approach includes what the KSW expansion considers to be subleading
terms.

The right plot of Fig.~\ref{fig:2yuk} has a comparison of the
NNLO result for the phase shift, as determined from
\eqsii{exp1}{exp2}, suggesting that the higher order corrections to
the 
expansion become important at a scale around $p\sim 400$~MeV. 
That the breakdown is
a factor of two smaller than for a delta-shell ``rho''
is in agreement with the analysis of the two-Yukawa model in
Ref.~\cite{steeleb}. 

Instead of using the actual value for $\wt r_0$, we could also treat it 
as a free parameter and use a one-parameter global fit to data.
This gives $\wt r_0=0.400$~fm at
NLO and $\wt r_0=0.671$~fm at NNLO. The deviations from the true value 
of $\wt r_0$ by $47\%$ and $12\%$ respectively are roughly what is
expected from the estimates 
$\order(\alpha_\pi/\alpha_{crit})$ and 
$\order([\alpha_\pi/\alpha_{crit}]^2)$,
where $\alpha_{crit}\simeq 0.25$ is the critical pion
coupling that would lead to a bound state in the OPE potential.

Note that a low-momentum fit will give a more accurate result at
low-momentum~\cite{RS}, but overall only differ from our approach by
higher order terms.
The result will be the same breakdown scale as we observe
in Fig.~\ref{fig:2yuk}.
However, it is important to realize at NNLO no new parameter is
required in our way of fitting, in contrast to the  method
employed in Ref.~\cite{RS}, so that our fit has one fewer free parameters. 

\subsection{The Three-Yukawa Model}
\label{sec:4c}

\begin{figure}
\begin{center}
\leavevmode
\hbox{
\epsfxsize=3in
\epsffile{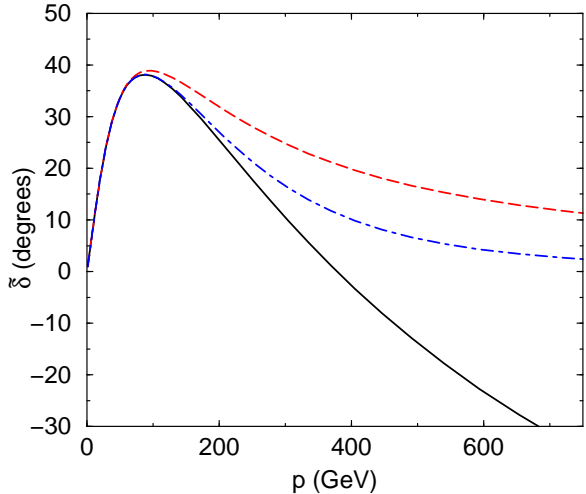}
}
\end{center}
\caption{\label{fig:3yuker}  The phase shift $\wt\delta$ due to
  scattering from the combined 
  $\rho$ and $\sigma$ contributions to the potential $V(r)$ in
  \eq{3yuks}. The exact result (solid) is 
  compared with the effective range expansion to order
  $p^2$ (dashed) and $p^4$ (dot-dashed). }
\end{figure}

The two-Yukawa model is unable 
to reproduce  the large  effective range
$r_0=2.63$~fm of the $\si$ channel for any choice of couplings in \eq{2yuk}.
So as a final example, we add one more Yukawa interaction in order to 
have the proper effective range.  We take for our potential
\beq
V=V_\pi + V_S\ ,\qquad V_\pi = - \alpha_\pi \frac{e^{-m_\pi r}}{r}\ ,\qquad 
      V_S= - \alpha_\sigma \frac{e^{-m_\sigma r}}{r} 
       + \alpha_\rho \frac{e^{-m_\rho r}}{r} \, ,
\eqn{3yuks}
\eeq
with the physical values for $m_\pi$, $m_\rho$, and $\alpha_\pi$,  and with
$m_\sigma=500\MeV$. We then determine the final two parameters in order 
to obtain $a=-23$~fm and $r_0=2.6$~fm, yielding  $\alpha_\sigma=7$ and
$\alpha_\rho=14.65$.
This three-Yukawa model is reminiscent of the Bonn
potential, which is known to model the data well.

Examining scattering from the short-range potential $V_S$ alone, we find
\beq
\wt a = -3.3~{\rm fm}\ ,\qquad \wt r_0 = 2.59~{\rm fm}\ ,
\eeq
which immediately suggests that we will encounter serious problems due
to the large size of $\wt r_0$.  Recall that $\wt r_0$ is assumed
to be $\order(1/\Lambda_S)$ in all effective field theory approaches
proposed to date.
For this model, however, the scale $\Lambda_S$ corresponds to
$m_\sigma/2=250\MeV$, whereas $\wt r_0 
\simeq 1/(80\MeV)$ corresponds to a much smaller scale.  

Performing an effective range expansion for $\wt \delta$,
we obtain the result shown in  Fig.~\ref{fig:3yuker}.
The short-distance effective range expansion, if carried to high 
enough order, does indeed have a radius of convergence set by $p\sim
m_\sigma/2$.  
However, the expansion at $\order(p^2)$ deviates from the exact result 
already at momenta below $m_\pi$.
This suggests that an EFT expansion to the same order 
will also fail at momenta below the pion mass, even with the pion
included explicitly.

\begin{figure}
\begin{center}
\leavevmode
\hbox{
\hspace{-.5cm}
\epsfxsize=3.4in
\epsffile{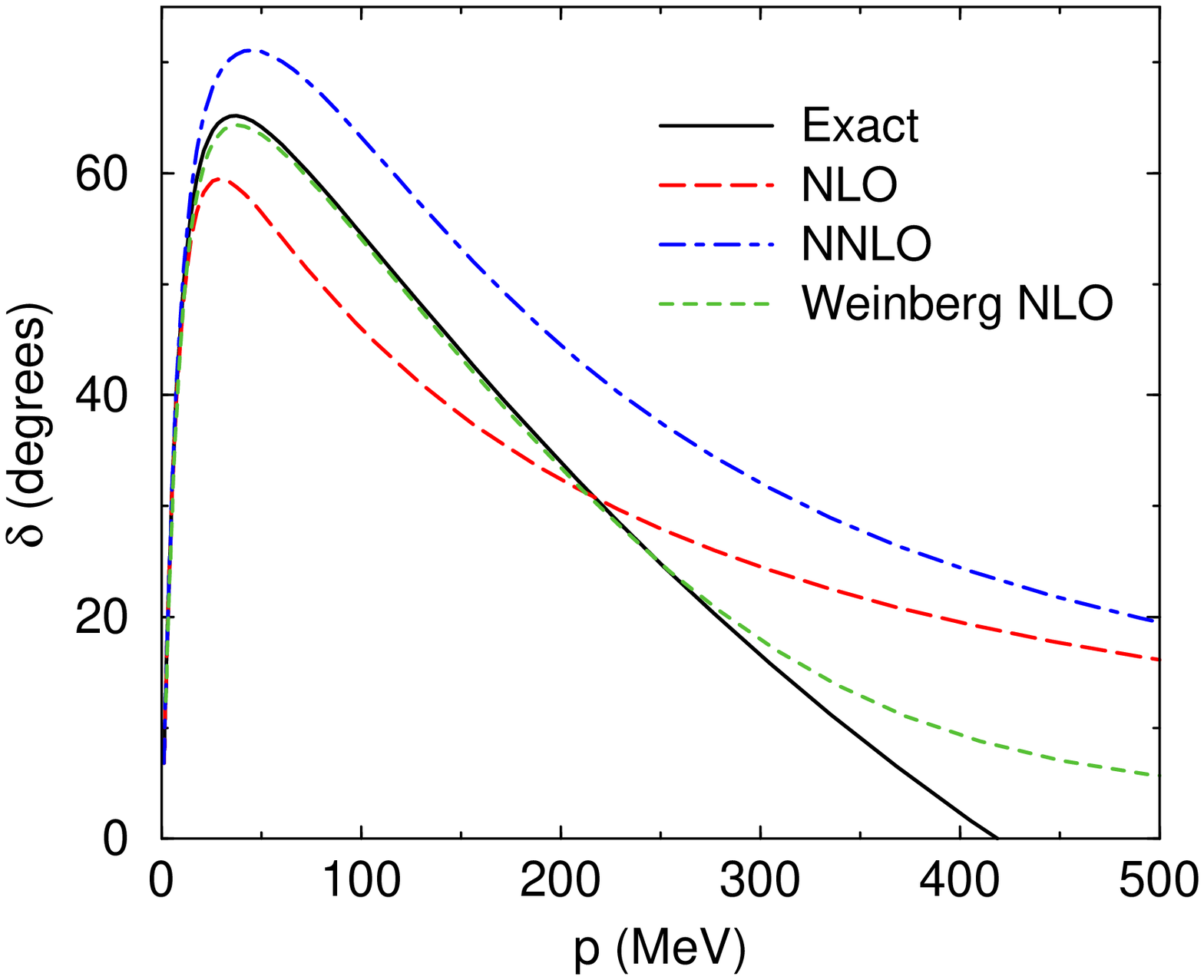}
\epsfxsize=3.4in
\epsffile{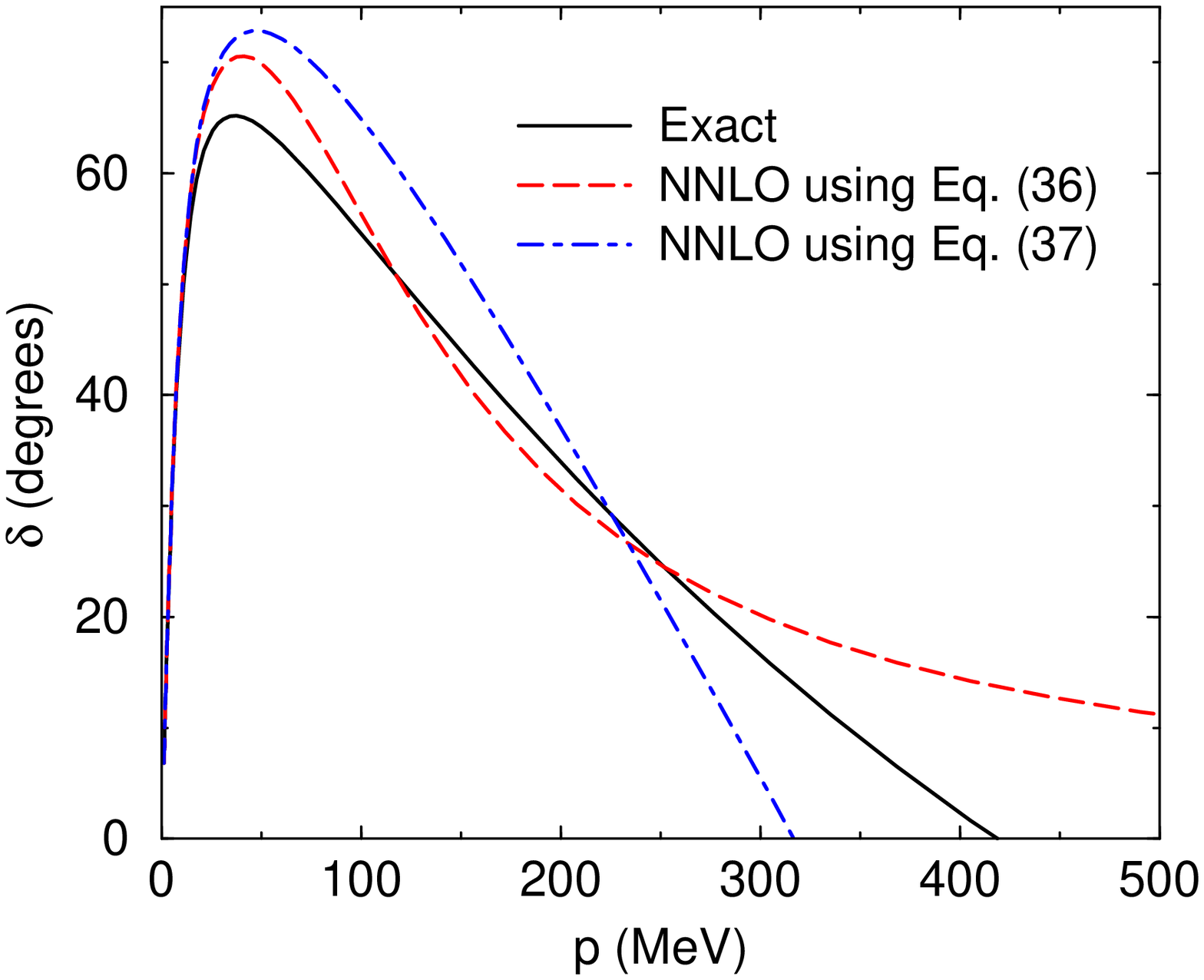}
}
\end{center}
\caption{\label{fig:3yuk}The algorithm of \S\ref{sec:3ciii} 
for the KSW expansion
applied to the three-Yukawa model. The right plot shows 
the NNLO results using the different expansion schemes
\eqsii{exp1}{exp2} to determine the
breakdown scale.
}
\end{figure}

Applying the EFT to the three-Yukawa model reveals this to be the case.
The results for NLO and NNLO using the exact $\wt r_0=2.59$~fm are
shown in Fig.~\ref{fig:3yuk}.
The EFT with a global fit instead gives $\wt r_0=1.68$~fm at NLO and
$\wt r_0=3.71$~fm at NNLO, also showing no signs of convergence.  
Similarly, the full effective range gives $r_0=3.7$~fm at NLO and
$r_0=1.1$~fm at NNLO, both equally far from the exact value $2.6$~fm.

The Weinberg approach result does better overall, only deviating from
the exact result around $300$~MeV.  This indicates that the subleading
order terms which are accounted for in this case are larger
than anticipated from the KSW power counting.
At best, the breakdown scale for the KSW power counting 
appears also to be around $300$~MeV as seen in the right plot of
Fig.~\ref{fig:3yuk}. 
If the $\sigma$ meson can be described accurately by non-irreducible
two-pion effects, the Weinberg counting would include this at
NLO and could possibly improve the radius of convergence even 
further~\cite{sfurn}.

\subsection{Lessons Learned}
\label{sec:4d}

The point of this section was to test the efficacy of the expansion
algorithm of \S\ref{sec:3ciii}.  An evident trend can be spotted:  the
larger the parameter $\wt r_0$, the slower the expansion
converges, verifying the claims of Cohen and Hansen \cite{CH3} and
discussed by Cohen  
\cite{Cohen:1999uh} that the short distance physics in $NN$ scattering
may not look sufficiently short.   
In the first model, with both long- and short-range
delta-shell potentials, $\wt r_0=0.36$~fm and the convergence is excellent.
The second example includes a realistic OPE potential and
an attractive short-range Yukawa, giving $\wt r_0=0.77$~fm.
Here too, the NNLO calculation does quite well, although not as good as 
in the first example.  The third example with three-Yukawa interactions 
has an enormous $\wt r_0 = 2.6$~fm, and the expansion does poorly at
NNLO. 
Note that in the EFT, the only difference between the two- and
three-Yukawa models is the value for $\wt r_0$. In particular, both
examples use the same  OPE potential for the long-range
interaction.  
This suggests that the  
claims in the literature that pions cannot be treated perturbatively 
\cite{Gegelia} 
are  at least oversimplified, as the perturbative expansion works well 
in the two-Yukawa example.

It is clear that the three-Yukawa example fails for several reasons.
First of all, $p \wt r_0 $ is assumed small at $p\sim m_\pi$,
whereas numerically it is large.  
Therefore, the fact that  $p \wt r_0$ is not
kept to all orders in the KSW expansion is a practical problem. A
possible resolution is to adopt the
method for summing the effects of large $r_0$ by means of a dibaryon 
field, as discussed in Refs.~\cite{DBK,KSW}.
Secondly, the effective range expansion plot in Fig.~\ref{fig:3yuker}
shows that, even if the effects of a large $\wt r_0$ are summed to all
orders, the EFT to NNLO may still not give accurate predictions
for the phase shift above $p\sim m_\pi$.  
This depends on the particular form of the short-distance 
physics and has nothing to do with the pion.
Finally,
the very nature of the three-Yukawa potential implies that even if one 
 sums the effects of $\wt r_0$ to all orders and pushes beyond
NNLO to include $\wt v_2$, the convergence will never extend to $p\sim
m_\rho$, but 
instead will fail at $p\sim m_\sigma/2 =250\MeV$.  This 
obstacle does not depend on how one chooses to perform the EFT
expansion, and can only be  surmounted by explicitly including the
$\sigma$ field in the EFT. If chiral symmetry cannot describe the
$\sigma$ as irreducible two-pion effects, 
an EFT including explicit $\sigma$
mesons would have little 
to offer beyond conventional modeling techniques for nuclear physics.

We have not included an analysis of the real data in this section,
since to date the full  
amplitude at NNLO has not been calculated.  In particular, the
contributions from radiation  pions (on-shell, propagating pions)
\cite{MS2}, has
not been computed to this order.  When these terms are known, then our 
algorithm can be readily applied to the full NNLO amplitude, testing
whether the data exhibits the same behavior as the three-Yukawa
model. Much of the difficulties encountered in our
three-Yukawa model would be alleviated if
radiation pions, which are absent in 
the potential models, are found to make a significant contribution to $r_0$;
however, we have no reason to
expect this to happen.

\section{Discussion}
\label{sec:5}

In this paper, we have examined the nature of the KSW
expansion in a systematic fashion by using 
models for potential scattering  between
``nucleons''.
These model potentials contain both short- and long-range
interactions and by construction give rise to a large scattering length, 
which can lead to complications in the implementation of
power counting methods. 
By beginning with an explicitly solvable toy model, we 
motivated an algorithm for fixing the unknown coupling constants of
the EFT, directly relating them to the effective range expansion for
scattering from the short-distance potential alone.  This procedure is
similar to the modified effective range expansion
\cite{modER1,steeleb}.  We then applied this algorithm to
three different models.

We find that for a realistic pion mass and coupling,
the quality of the expansion is good when the relation
$m_\pi \wt r_0 < 1$ is valid, where $\wt r_0$ is the effective range due to
the short-range part of the potential alone.  An exception is seen at
low momentum, where the strict perturbative pion expansion fails due to the
delicate interplay between short-range and pion 
interactions that gives rise to a large scattering length.  This
problem is dealt with by reordering the expansion to correctly account 
for the scattering length or pole in the amplitude at LO, as discussed 
previously in Refs.~\cite{MS1,RS}. The algorithm we provide for 
fixing the contact interactions of the EFT correctly account
for this reordering. 

It remains to be seen, however, if $\wt r_0$ 
needs to be large for real $\si$ $NN$ scattering, 
or whether the large effective range $r_0$ gets sizable contributions from 
radiation pion
effects. In potential models, however, 
$\wt r_0$ does have to be large and results in the KSW expansion doing
poorly, a discovery made previously in Ref.~\cite{CH3}.  In this case, the
Weinberg expansion, which sums $\wh C_2$ (and 
hence $\wt r_0$) to  
all orders, may do better.  However, as the Weinberg procedure involves 
summing the extended pion interaction to all orders at LO in the EFT
expansion, it seems  to offer only modest
benefits over conventional potential model techniques for systems
with more than two nucleons.  In order to 
sum $\wt r_0$ to all orders while still treating pions perturbatively, 
it may be advantageous to make use of the dibaryon field as discussed
in \cite{DBK,KSW}, and used to advantage in the three-body problem
\cite{Bvk,geg3b}. 
A revised power counting scheme, perhaps
treating $\wt r_0\sim 1/m_\pi$, remains to be worked out. 


\vskip1truein
\centerline{\bf Acknowledgements}
\medskip
D.K.\ would like to thank T.\ Mehen, G.\ Rupak, N.\ Shoresh, I.\
Stewart, and M.\ Wise  
for useful discussions, and is
supported in part by the U.S.\ Dept.\ of Energy under
Grants No.\ DOE-ER-40561 and No.\ DE-FG03-92ER40689;  
he thanks the Santa Cruz Institute for
Particle Physics and the TH Division of CERN for their hospitality 
during various stages of this work.
J.S.\ would like to thank R.\ Furnstahl for useful discussions
and is supported by the National Science Foundation under Grants
No.~PHY-9511923 and PHY-9258270.

\appendix
\section{Summing Up Toy Pion Graphs}
\label{sec:6}

The toy pions of \eq{vldef} have a simple enough form to allow an analytic
summation of all contributions.
Following the discussion in Ref.~\cite{KSWa}, there are three classes
of graphs to compute: i) pion ladders, 
ii) a single vertex of the short-distance
potential dressed by pion exchange on one side, and 
iii) two vertices of
the short-distance potential with pion exchange in between.

\begin{figure}[h]
\epsffile[50 690 417 720]{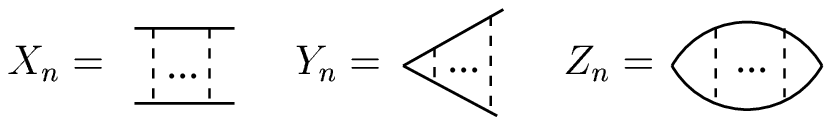}
\end{figure}

\noindent
They take the form
\beq
X_n &\equiv& \bra{{\bf p}}\hat V_L (\hat G_E \hat V_L)^n \ket{{\bf p'}}
 = -{4\pi\over M m_\pi}
\({g_\pi \sin^2\!\xi\over  \xi^2}\)\({g_\pi e^{i\xi} \sin \xi\over
 \xi}\)^n \, ,
\\
Y_n &\equiv&\bra{{\bf r}=0}  (\hat G_E \hat V_L)^n \ket{{\bf p'}} =
\({g_\pi e^{i\xi}\sin \xi \over
 \xi}\)^n\, , 
\\
Z_n &\equiv& \bra{{\bf r}=0}\hat  G_E \hat V_L \hat G_E (\hat V_L
\hat G_E)^n \ket{{\bf 
r'}=0}=
 -{Mm_\pi\over 4\pi}\({g_\pi e^{2 i \xi} }\)\({g_\pi e^{i\xi}\sin \xi
\over  \xi}\)^n  \, ,
\eeq
where $p^2=(p')^2=ME$, $\xi=p/m_\pi$,  and we have projected onto the
$s$-wave. 

These graphs are trivially summed as they form a geometric series.  
Using the PDS subtraction scheme, the
final result for the exact Feynman amplitude in the effective theory
with pions is
\beq
 \CA^{{\rm eff}} &&= -\sum_{n=0}^\infty X_n -{\frac{4\pi}{M} \wh C(p^2;\mu)
\[\sum_{n=0}^\infty Y_n\]^2\over 1- \wh C(p^2;\mu) \[-\mu-ip +
\frac{4\pi}{M} \sum_{n=0}^\infty Z_n\] } \nonumber \\ 
\nonumber\\
&&= \CA_L  +  {4\pi/M\over f(\xi)^2  \[ 
-\mu-ip -\wh C^{-1}(p^2;\mu) \] - g_\pi m_\pi  e^{2i\xi} f(\xi)}\ ,
\eqn{ampfulleftA}
\eeq
where we have defined $f(\xi)$ and $\CA_L$ as in \eqsii{fdef}{amppi}.

\section{Toy Pion Amplitude to NNLO}
\label{sec:7}

Taking the exact EFT result for the toy pions
\eq{ampfulleft} and expanding in powers of $Q$ according to the
rules of \S\ref{sec:2c} and \S\ref{sec:3ciii}, the first few terms 
are
\beq
\CA_{-1} = 
-{\frac{4\pi}{M} \wh\Cs{0}{-1}\over 1 + \wh \Cs{0}{-1} (\mu + i p)}\, ,
\eeq
\beq
\CA_0 = (\wh \Cs{2}{-2} p^2 + \wh\Cs{0}{0}) \frac{d\CA_{-1}}{d\wh\Cs{0}{-1}} 
-X_0 + 2\CA_{-1} Y_1 - \CA_{-1}^2 Z_0\ ,
\eeq
\beq\begin{array}{rcl}
\CA_1 &=& \displaystyle
\(\wh\Cs{4}{-3} p^4 + \wh\Cs{2}{-1} p^2  + \wh\Cs{0}{1}\)
\frac{d\CA_{-1}}{d\wh\Cs{0}{-1}} 
+\frac12 \(\wh\Cs{2}{-2} p^2 + \wh\Cs{0}{0}\)^2
\frac{d^2\CA_{-1}}{d\wh\Cs{0}{-1}{}^2}
 \\  \\&&
\displaystyle
+ \(\wh\Cs{2}{-2} p^2 + \wh\Cs{0}{0}\) \frac{d}{d\wh\Cs{0}{-1}} \( 2 \CA_{-1}
Y_1 - \CA_{-1}^2 Z_0 \)
\\ \\&& 
- X_1 +\CA_{-1}\(Y_1^2 + 2 Y_2\)- \CA_{-1}^2 \( 2 Y_1 Z_0 + Z_1 \)
+\CA_{-1}^3 Z_0^2\, .
\end{array}
\eeq
Using the values for the $X_n$, $Y_n$, and $Z_n$ in
Appendix~\ref{sec:6}, these expressions can be rewritten as
\beq
(p\cot\delta)_{\rm LO} &=& -\frac1{a}\, ,
\\
(p\cot\delta)_{\rm NLO} &=& (p\cot\delta)_{\rm LO} +
\frac{(m_\pi-1/a)^2}{\Lambda_\pi} +
\frac{2p^2}{3m_\rho} - \frac{m_\pi^2}{\Lambda_\pi} \left(
\cos \xi - \frac{\sin \xi}{pa}  \right)^2\, ,
\\
(p\cot\delta)_{\rm NNLO} &=& (p\cot\delta)_{\rm NLO} +
\frac{( m_\pi - 1/a)^3}{\Lambda_\pi^2} 
- \frac{p^2}{3a m_\rho^2} + \frac{m_\pi^3}{\Lambda_\pi^2} 
\frac{\sin \xi}{\xi} \left( \cos \xi- \frac{\sin\xi}{pa} \right)^3 
\nonumber\\
&&{}- \frac{2m_\pi}{\Lambda_\pi} \frac{\sin \xi }{\xi}
\left( \frac{(m_\pi-1/a)^2}{\Lambda_\pi} + \frac{2p^2}{3m_\rho}
\right) 
\left( \cos \xi - \frac{\sin \xi}{pa} \right)\, . 
\eeq


\end{document}